\documentclass{v16nufact}

\def\be{\begin{equation}}
\def\ee{\end{equation}}
\def\bea{\begin{eqnarray}}
\def\eea{\end{eqnarray}}

\newcommand{\Photo}{}

\begin{document}
\vspace*{4cm}
\title{SHOULD T2K RUN IN DOMINANT NEUTRINO MODE TO DETECT CP VIOLATION ?}

\author{MONOJIT GHOSH}

\address{Department of Physics, Tokyo Metropolitan University, Hachioji, Tokyo 192-0397,  Japan}

\maketitle\abstracts{
The main aim of the T2K experiment in Japan is to discover CP violation in the leptonic sector by measuring the Dirac phase $\delta_{CP}$. For that purpose T2K has already started collecting data
in both neutrino and antineutrino mode. But in this work we will show that, in T2K the main role of the antineutrinos is to resolve the octant degeneracy. 
If the octant is known then the pure neutrino run of T2K is capable to give the maximum CP sensitivity.
On the otherhand in the experiment like NO$\nu$A, antineutrinos are still useful even when octant is known. Thus we propose that let T2K run in the dominant neutrino mode whereas the antineutrino component of
the other experiments can resolve the octant degeneracy in T2K. As an example we show that if T2K is combined with the experiments NO$\nu$A and ICAL@INO, then T2K will have the potential to discover CP violation
with maximum sensitivity in the dominant neutrino mode.
}

\section{Introduction}

Neutrino oscillation is a quantum mechanical interference phenomena in which neutrinos with one flavour evolve into another flavour over microscopical distance and time. In standard
three flavour scenario, mathematically neutrino oscillations can be described by three mixing angles ($\theta_{12}$, $\theta_{13}$, $\theta_{23}$), two mass squared differences ($\Delta m_{21}^2$, $\Delta m_{31}^2$) and one Dirac type
CP phase $\delta_{CP}$. The measurement of $\theta_{12}$ and $\Delta m_{21}^2$ comes from solar and KamLAND data \cite{Eguchi:2002dm} whereas the parameters $\Delta m_{21}^2$ and $|\Delta m_{31}^2|$ 
are measured by the accelerator \cite{Adamson:2011qu}
and atmospheric neutrino experiments \cite{Hirata:1988uy}. 
The The smallest mixing angle $\theta_{13}$ has been measured quite recently by the short-baseline reactor experiments \cite{An:2012eh}. At present the task of the current /next generation 
experiments is to measure the following unknown parameters: (i) neutrino mass hierarchy i.e, $\Delta m_{31}^2 > 0$ so called normal hierarchy (NH) or $\Delta m_{31}^2 < 0$ so called inverted hierarchy (IH), 
(ii) the octant of the mixing angle $\theta_{23}$ i.e., $\theta_{23} <45^\circ$ so called lower octant (LO) or $\theta_{23} > 45^\circ$ so called higher octant (HO) and 
(iii) the value of the Dirac CP phase $\delta_{CP}$. In this paper we will study the role of antineutrinos in T2K to discover CP violation in the leptonic sector.

\section{The T2K experiment}

T2K is a long-baseline neutrino oscillation in Japan \cite{Abe:2015awa}. 
In this experiment muon neutrinos which are produced in the JPARC facility are detected at Kamioka after traveling a distance of 295 km. This experiment has already seen 
events in both neutrino and antineutrino mode of running and now it is collecting more data to establish the CP violation in the leptonic sector on a firm footing. The CP sensitivity of the T2K experiment comes 
from the electron neutrino appearance channel given by:
\begin{eqnarray}
P(\nu_\mu \rightarrow \nu_e) = P_{\mu e} & = 4 s^{2}_{13}s^{2}_{23}\frac{\sin^{2} (A-1)\Delta}{(A-1)^2}
\nonumber + \alpha^{2} \cos^{2}\theta_{23} \sin^{2}2 \theta_{12} \frac{\sin^{2} A\Delta}{A^2} \\ & +\alpha s_{13} \sin 2\theta_{12}  \sin 2\theta_{23}\cos(\Delta+\delta_{cp}) \frac{\sin (A-1)\Delta}{(A-1)}\frac{\sin A\Delta}{A} \label{p_mu_e}
\end{eqnarray}
where $ s_{ij}(c_{ij})=\sin \theta_{ij}(\cos \theta_{ij}) $, $ \alpha = \Delta m^{2}_{21}/ \Delta m^{2}_{31} $, $ \Delta =  \Delta m^{2}_{31} L / 4 E $,
and $ A = 2E V/ \Delta m^{2}_{31}$, where $ V(x) \simeq \pm 7.56 \times 10^{-4} \left( \frac{\rho (x)}{g/cc}\right) Y_{e}(x) $ eV is the Wolfenstein matter term. The above equation is for neutrinos.
The probability of the antineutrinos can be obtained by replacing $A \rightarrow -A$ and $\delta_{CP} \rightarrow \delta_{CP}$. 
As the sign of $\delta_{CP}$ is opposite in neutrinos and antineutrinos, we understand that it is very important to have data from antineutrino run of T2K to establish CP violation 
in the leptonic sector on a firm
footing. But apart from that one also needs to understand how antineutrinos help in the improvement of CP sensitivity. We know that CP sensitivity of T2K is suffered by parameter degeneracy \cite{Ghosh:2015ena}.
In parameter degeneracy, two sets of oscillation parameter give rise to equal value in the neutrino oscillation probability which makes it difficult to determine neutrino oscillation parameters uniquely.
In recent data, there are two types of parameter degeneracy: (a) hierarchy-$\delta_{CP}$ degeneracy \cite{Prakash:2012az} and (ii) octant-$\delta_{CP}$ degeneracy \cite{Agarwalla:2013ju}. 
It is well known that hierarchy-$\delta_{CP}$ degeneracy behaves
similarly in neutrinos and antineutrinos but the dependence of octant-$\delta_{CP}$ degeneracy in neutrinos is different than antineutrinos. Thus we expect that antineutrinos will play a very important role in
resolving the octant-$\delta_{CP}$ degeneracy. In the next section we will study how antineutrinos help in the CP sensitivity by ruling out the wrong octant solutions.

\section{Sensitivity of T2K}

The CP violation (CPV) discovery $\chi^2$ is defined as the capability of an experiment to distinguish a true value of $\delta_{CP}$ other than $0^\circ$ and $180^\circ$.
\begin{figure}
\includegraphics[width=0.5\linewidth]{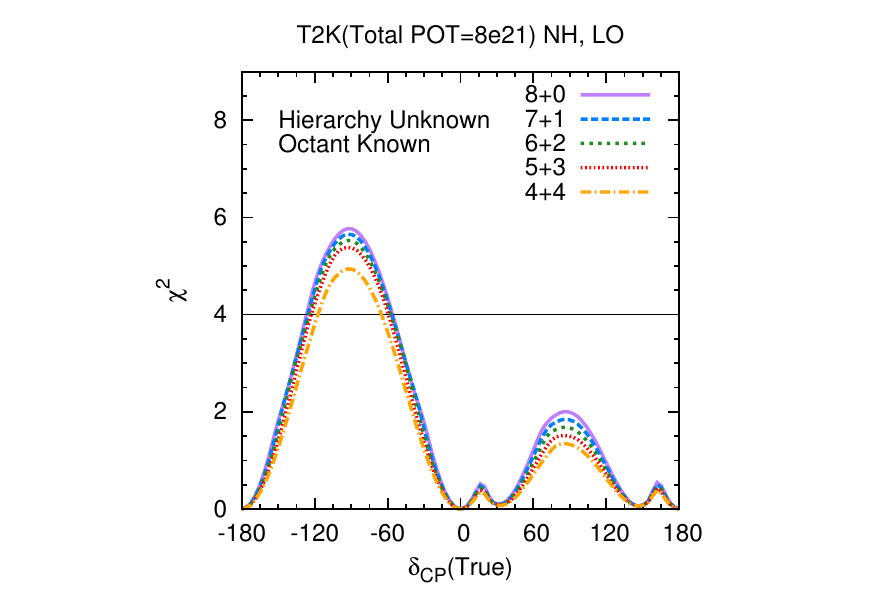}
\hspace{-0.9 in}
\includegraphics[width=0.5\linewidth]{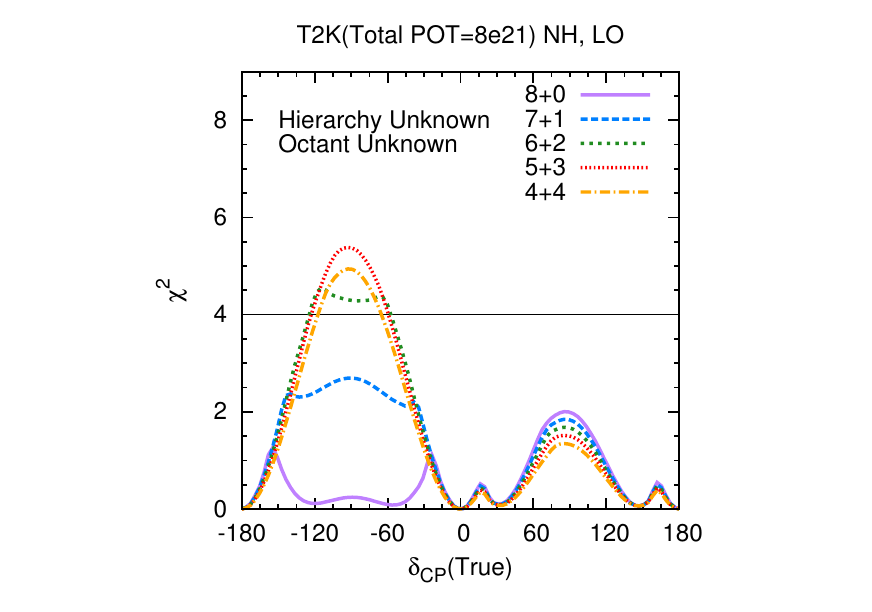} \\
\includegraphics[width=0.5\linewidth]{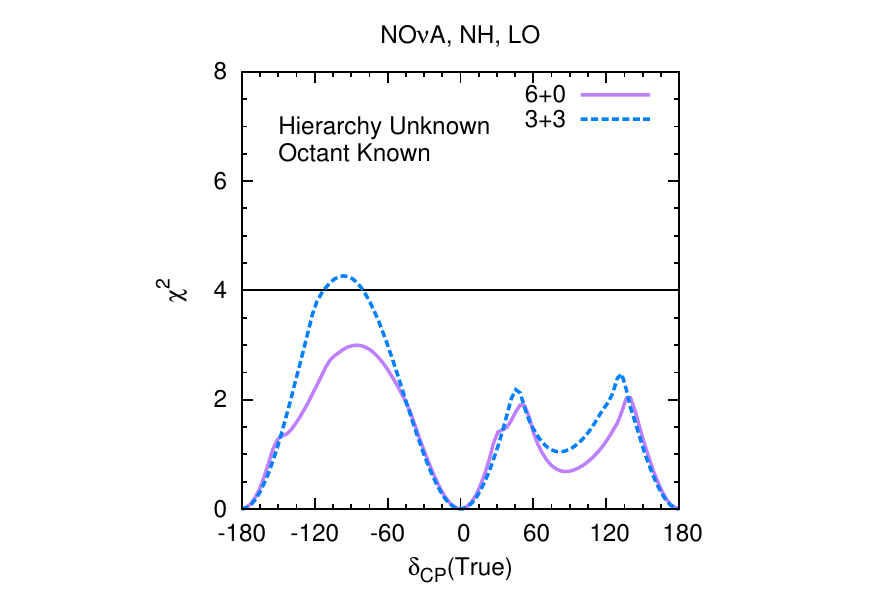}
\hspace{-0.9 in}
\includegraphics[width=0.5\linewidth]{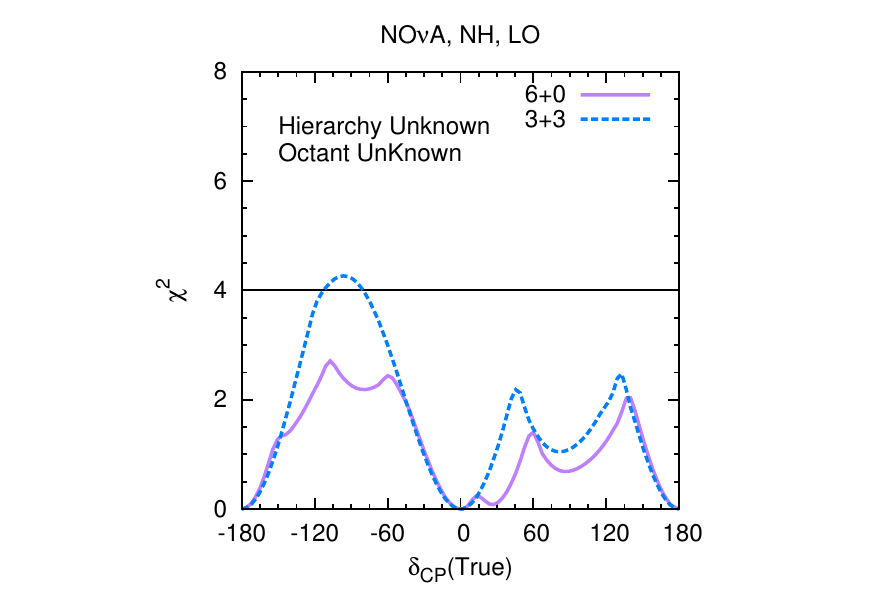}
\caption{CP Sensitivity of T2K and NO$\nu$A for NH-LO.}
\label{fig1}
\end{figure}
In the upper panels of Fig. \ref{fig1} we have plotted the CPV discovery $\chi^2$ of T2K for a total exposure of $8 \times 10^{21}$ protons on target (pot). 
We have divided this exposure among neutrinos and antineutrinos
in units of $10^{21}$ pot. An exposure 8+0 means, T2K runs in pure neutrino mode. Fig. \ref{fig1} is for NH-LO i.e., $\Delta m_{31}^2 = +2.4 \times 10^{-3}$ and $\theta_{23}=39^\circ$. In the left panel octant 
is assumed to be known and in the right panel octant is unknown. From the plots we notice that antineutrinos helps in the CP sensitivity if octant is unknown. However if octant is known then adding antineutrino run
causes a decrease in the sensitivity. This is because replacing neutrinos by antineutrinos reduces the statistics in a significant way.
But this is not the case for NO$\nu$A. NO$\nu$A is another long-baseline experiment in Fermilab having a baseline of 812 km \cite{Adamson:2016xxw}. 
In the lower panels of Fig. \ref{fig1} we see that
for NO$\nu$A antineutrinos help in improving the CP sensitivity even when octant is unknown. 
This lies in the fact that for T2K the flux and oscillation maxima peaks at the same energy but for NO$\nu$A, the flux and oscillation maxima corresponds
to different energy.
\begin{figure}
\includegraphics[width=1.0\linewidth]{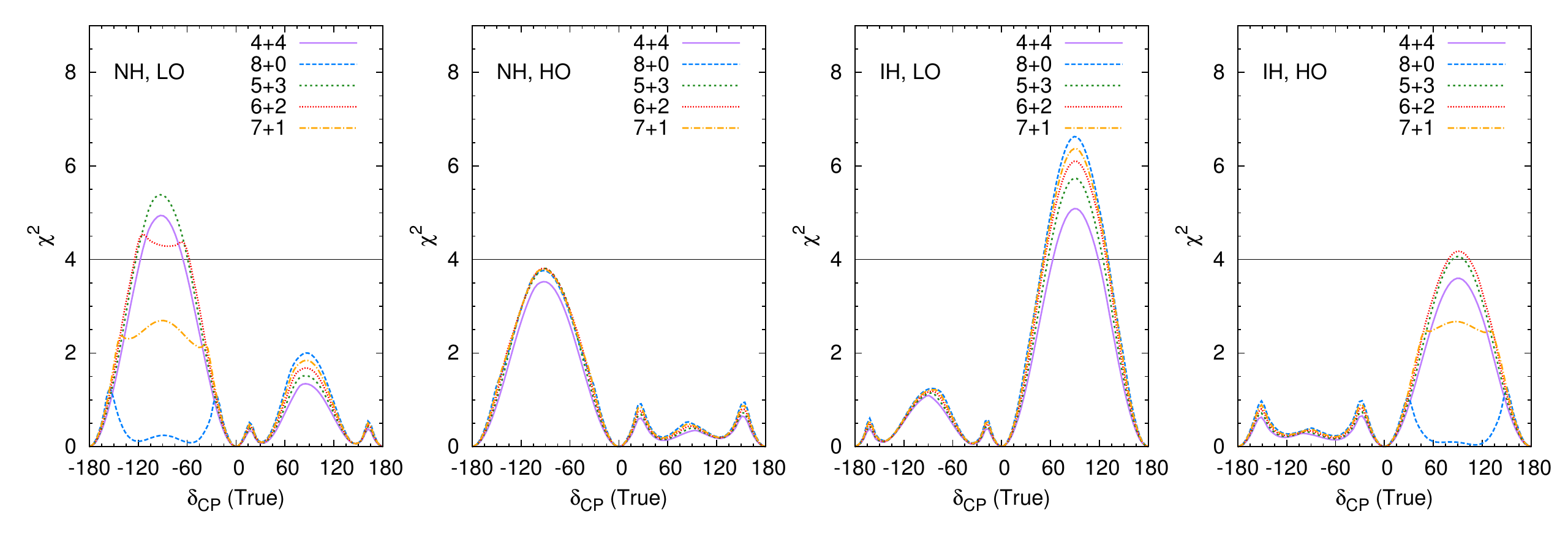}
\caption{CP Sensitivity of T2K for all the four combinations of hierarchy and octant.}
\label{fig2}
\end{figure}
In Fig. \ref{fig2}, we have plotted the same as that of Fig. \ref{fig1}, but for all the four combinations of hierarchy and octant. Here IH corresponds to $\Delta m_{31}^2 = -2.4 \times 10^{-3}$ and
HO corresponds to $\theta_{23}=51^\circ$. From the plot we see that antineutrino helps only for $-90^\circ$-NH-LO and $+90^\circ$-IH-HO. Thus we conclude for T2K, antineutrinos help only in a limited parameter
space. To overcome this problem we suggest that let T2K run in the dominant neutrino mode whereas antineutrinos from the other experiments can compensate the antineutrino runs of T2K. In the next section we discuss 
the combined CP sensitivity of the T2K, NO$\nu$A along with the atmospheric neutrino experiment ICAL@INO. ICAL is a proposed experiment in India which will use a 50 kt magnetized iron calorimeter detector 
to study oscillation of the neutrinos coming from Earth's atmosphere \cite{Ahmed:2015jtv}.

\section{CP sensitivity of T2K, NO$\nu$A and ICAL}
\begin{figure}
\includegraphics[width=1.0\linewidth]{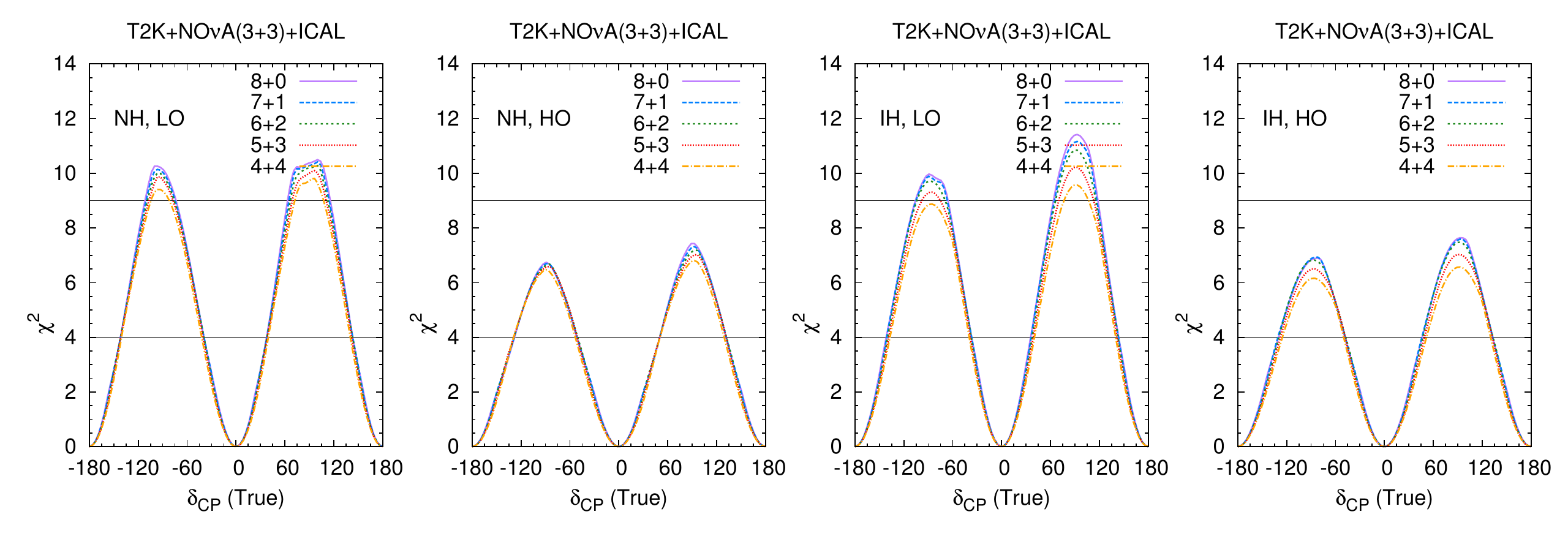}
\caption{CP Sensitivity of T2K, NO$\nu$A and ICAL for all the four combinations of hierarchy and octant.}
\label{fig3}
\end{figure}

In Fig. \ref{fig3}, we have plotted the combined CP sensitivity of T2K, NO$\nu$A and ICAL for all the four true combinations of hierarchy and octant. From the plots we see that when these experiments are added to 
T2K data, the best CP sensitivity of T2K comes from the 7+1 combination for every true combination of hierarchy and octant.

\section{Conclusion}

In this paper we have studied the role of antineutrinos in T2K to discover CP violation in the leptonic sector. We have shown that in T2K the main role of the antineutrinos is to remove the wrong octant solutions.
For the parameter space where there is no wrong octant solution, pure neutrino run of T2K gives the best CP sensitivity. The scenario is different for NO$\nu$A. For NO$\nu$A, due to some additional synergy
addition of antineutrinos help in the CP sensitivity even when octant is known. To overcome this problem of T2K we propose that let T2K run in the dominant neutrino mode while the antineutrinos from other experiments
can compensate the antineutrino runs of T2K. We have demonstrated this for the experiment NO$\nu$A and ICAL. In our analysis we find that when data from NO$\nu$A and ICAL is added to T2K, best CP 
sensitivity can be obtained from T2K if it runs in dominant neutrino mode. 
Though we have shown our results for $\theta_{23}=39^\circ$ and $51^\circ$, our conclusion remains same for all the other values of $\theta_{23}$.
The  The results obtained in this work are important to design the future runs of T2K. 
For more details we refer to \cite{Ghosh:2015tan} on which this work is based upon.

\section*{Acknowledgements}
This  work is supported by the “Grant-in-Aid for Scientific
Research of the Ministry of Education, Science and Culture,
Japan”, under Grant No.  25105009.


\section*{References}

\begin{thebibliography}{99}
\bibitem{Eguchi:2002dm} 
  K.~Eguchi {\it et al.} [KamLAND Collaboration],
  Phys.\ Rev.\ Lett.\  {\bf 90}, 021802 (2003)
  [hep-ex/0212021].
  
\bibitem{Adamson:2011qu} 
  P.~Adamson {\it et al.} [MINOS Collaboration],
  Phys.\ Rev.\ Lett.\  {\bf 107}, 181802 (2011)
  [arXiv:1108.0015 [hep-ex]].
  
\bibitem{Hirata:1988uy} 
  K.~S.~Hirata {\it et al.} [Kamiokande-II Collaboration],
  Phys.\ Lett.\ B {\bf 205}, 416 (1988).
  
\bibitem{An:2012eh} 
  F.~P.~An {\it et al.} [Daya Bay Collaboration],
  Phys.\ Rev.\ Lett.\  {\bf 108}, 171803 (2012)
  [arXiv:1203.1669 [hep-ex]].
  
\bibitem{Abe:2015awa} 
  K.~Abe {\it et al.} [T2K Collaboration],
  Phys.\ Rev.\ D {\bf 91}, no. 7, 072010 (2015)
  doi:10.1103/PhysRevD.91.072010
  [arXiv:1502.01550 [hep-ex]].

  \bibitem{Ghosh:2015ena} 
  M.~Ghosh, P.~Ghoshal, S.~Goswami, N.~Nath and S.~K.~Raut,
  Phys.\ Rev.\ D {\bf 93}, no. 1, 013013 (2016)
  [arXiv:1504.06283 [hep-ph]].

\bibitem{Prakash:2012az} 
  S.~Prakash, S.~K.~Raut and S.~U.~Sankar,
  Phys.\ Rev.\ D {\bf 86}, 033012 (2012)
  [arXiv:1201.6485 [hep-ph]].
  
\bibitem{Agarwalla:2013ju} 
  S.~K.~Agarwalla, S.~Prakash and S.~U.~Sankar,
  JHEP {\bf 1307}, 131 (2013)
  [arXiv:1301.2574 [hep-ph]].
  
\bibitem{Adamson:2016xxw} 
  P.~Adamson {\it et al.} [NOvA Collaboration],
  Phys.\ Rev.\ D {\bf 93}, no. 5, 051104 (2016)
  doi:10.1103/PhysRevD.93.051104
  [arXiv:1601.05037 [hep-ex]].
  
\bibitem{Ahmed:2015jtv} 
  S.~Ahmed {\it et al.} [ICAL Collaboration],
  arXiv:1505.07380 [physics.ins-det].

  
\bibitem{Ghosh:2015tan} 
  M.~Ghosh,
  Phys.\ Rev.\ D {\bf 93}, no. 7, 073003 (2016)
  [arXiv:1512.02226 [hep-ph]].




\end{thebibliography}
\end{document}